\documentstyle[preprint,prb,aps]{revtex}

\begin{document}
\title{Effect of LaAlO$_3$ Surface Topography on RF Current Distribution in Superconducting Microwave Devices}

\author{A. P. Zhuravel,$^{1}$ A. V. Ustinov$^{2}$, H. Harshavardhan$^{3}$, and Steven M. Anlage$^{4}$}
\address{$^{1}$B. I. Verkin Institute for Low Temperature Physics \& Engineering, National Academy of Sciences of Ukraine, 61164 Kharkov, Ukraine }
\address{$^{2}$Physics Institute III, University of Erlangen-Nuremberg, D-91058, Erlangen, Germany }
\address{$^{3}$Neocera, Inc., 10000 Virginia Manor Road, Suite 300, Beltsville, MD 20705 USA}
\address{$^{4}$Center for Superconductivity Research, and Materials Research Science and Engineering Center, Department of Physics, University of Maryland, College Park, MD 20742-4111 USA}
\maketitle

\begin{abstract}
A laser scanning microscope with a thermal spot size of about 4 $\mu$m is used to measure a quantity proportional to the rf current density in an operating superconducting co-planar waveguide microwave resonator.  The twinning of the LaAlO$_3$ substrate produces a meandering of the current at the edges due to irregularities in the wet etching of the YBa$_2$Cu$_3$O$_{7-\delta}$ film associated with substrate twin domain blocks, and a ($\sim$ 20\%) enhancement of the rf photoresponse at these locations.  These irregularities are candidates for enhanced nonlinear response from the device.  The effects of substrate twinning and the resulting edge features on the superconducting film are discussed and analyzed.
\end{abstract}


Compact high-temperature superconducting microwave filters made from thin film resonant devices still suffer from significant intermodulation distortion.\cite{Hein Thesis}  The source of nonlinearity in many of these devices is the large current buildup experienced at the edges of the patterned films.\cite{Hein Thesis,Brandt,van Duzer}  The nonlinear response is dominated by the edges because (for example) the third-order nonlinear response scales with the sixth power of the current.\cite{Vopilkin}  Although edge-current-free resonator designs have been demonstrated \cite{Chaloupka} and analyzed for their nonlinear properties,\cite{Dahm Disk} they may be too large in size for use in current applications.  

	LaAlO$_3$ (LAO) is a commonly employed substrate for high-temperature superconducting (HTS) thin films used in rf applications.  It has a modest dielectric constant of approximately 23.6 at 77 K and microwave frequencies \cite{Klein LAO} and is lattice-matched and chemically compatible with many cuprate superconducting materials.  LAO is cubic at the deposition temperature ($\sim$800$^0$C) of HTS films.  However it undergoes a cubic to rhombohedral transition as it is cooled below about 544$^0$C.\cite{Bueble}  The substrates commonly twin and produce a surface corrugation that can be more than 25 nm in height on lateral length scales of 1-10 $\mu$m.\cite{Sum}  The HTS films also generally twin at some temperature between the deposition temperature and T$_c$.  The effects of these two types of twinning on the rf properties of the HTS films have not been extensively investigated.\cite{Kastner} In particular the changes in edge current distribution caused by twinning, so crucial for determining the nonlinear properties of the device, have not been investigated.

	Microscopic techniques have been employed in the past to image rf currents in superconducting microwave devices with some success. \cite{C+N,Kaiser,Hu}   Culbertson and Newman imaged rf current distributions in superconducting resonators, but failed to see an expected re-distribution of currents upon increased rf power.  They also observed qualitatively that there was a correlation between the length scales of rf current variations and the granularity of their HTS film.\cite{C+N}  Other efforts to image rf current and field distributions in superconducting resonators have been done with relatively low spatial resolution.\cite{Kaiser,Hu}

	Important clues concerning the connection between edge-current buildup and nonlinear response come from investigation of geometry-dependent properties of transmission line devices.  The geometry-dependence of the nonlinear response of superconducting co-planar waveguides (CPW) has been extensively investigated by Booth, $et$ $al.$ \cite{Booth}  They find that narrower, shorter lines of smaller thickness tend to have the greatest nonlinear signal generation.  Dahm and Scalapino have demonstrated theoretically that acute edges on strip conductors can significantly enhance the local nonlinear response. \cite{Dahm Edge}  There is clearly a need to further investigate the microscopic details of the geometry dependence of nonlinear properties of superconducting transmission line devices.

 	The above experiments and theory demonstrate that the edges of the films are very important for understanding the linear and nonlinear properties of the device.  In this paper we study the effects of substrate twinning on the rf current distribution in wet etched superconducting microwave devices, based on microscopic imaging of the rf currents.

	A laser scanning microscope (LSM) is used to image the rf currents in a superconducting CPW resonator.  The LSM is schematically shown in Fig. \ref{Schematic}.  A laser supplies about 1 mW optical power at 670 nm light wavelength to the sample, and the beam is focused to a spot about 1.5 $\mu$m in diameter.  The intensity of the laser is modulated at a typical frequency of 100 kHz, producing a thermal spot size of about 4 $\mu$m in diameter.\cite{Zhuravel ASC2002}  The sample is a Ag-doped YBa$_2$Cu$_3$O$_{7-\delta}$ film with thickness t = 240 $\pm$ 28 nm, deposited by pulsed laser deposition. The LAO substrate is 500 $\mu$m thick, and the film is wet-etch patterned with dilute phosphoric acid into a CPW resonator with strip width of 500 $\mu$m, gap width of 650 $\mu$m, strip length of 7.75 mm, capacitively coupled through 500 $\mu$m gaps, and mounted in a brass microwave package.  The resonant frequency is about 5.2 GHz with a Q $\sim$ 2500 at a temperature of 78 K.

	The LSM develops contrast from the variations of local rf current density squared, J$_{rf}^2$.  It is commonly assumed that changes in the kinetic inductance of the device dominate the response of the resonator to the laser beam. \cite{C+N,Tsindlekht}  By applying a fixed frequency microwave signal at the inflection point f$_0$ of the $\mid$S$_{12}$(f)$\mid$$^2$ power transmission curve of the resonator (Fig. \ref{Schematic}, inset), and measuring changes in $\mid$S$_{12}$(f$_0$)$\mid$$^2$ as a function of position (x,y) of the laser beam perturbation on the sample, we image a quantity proportional to $\lambda ^2$(x,y) J$_{rf}^2$(x,y) $\delta \lambda$ .  Here $\lambda$(x,y) is the local value of the magnetic penetration depth and $\delta \lambda$ is the change in penetration depth caused by the laser heating. \cite{C+N}  Hence the LSM photoresponse is, to first order, an image of the rf current distribution squared as a function of laser perturbation position.  As shown in Fig. \ref{Schematic}, the changes in transmission measured at the modulation frequency of the laser (henceforth called the rf photoresponse), are detected as the laser beam is scanned over the surface of the sample by means of mirrors.  The optical reflectance of the material is simultaneously collected for comparison with the rf photoresponse image.

	Figure \ref{Fig2} demonstrates that the LSM images a quantity proportional to the square of the rf current distribution.  The image of the CPW strip near the fundamental resonance tone clearly shows the current bunching at the conductor edges.  The cosinusoidal standing wave pattern of the fundamental resonant mode is also evident.  A longitudinal line cut near the edge can be fit to the form cos$^2(kx+\phi)$, with $k$ = 0.39 mm$^{-1}$ and $\phi$ =4.62 radians, shown in Fig. \ref{Fig2}.  A simple estimate for the propagation wavenumber $k$ based on an the properties of an infinite transmission line with an effective dielectric constant \cite{Simons} of 15.1 yields $k$ = 0.42 mm$^{-1}$, a bit larger than the fit value, as expected for a CPW resonator with capacitive coupling.  Fits to other functional forms, such as cos$(kx+\phi)$, were not successful.  

	With a satisfactory understanding of the image contrast, we can now examine the detailed current profile along the edges of the CPW strip conductor.  Fig. \ref{Fig3} shows a magnified view of the edge in a 100 $\mu$m x 100 $\mu$m area.  Fig. \ref{Fig3}(a) shows the reflectivity of the film (top) and substrate (bottom).  The substrate shows the linear patterns of twin domain blocks running almost perpendicular to the patterned edge of the YBCO thin film.  The patterned edge is clearly crenelated on the length scale of the twin domain blocks, possibly due to inhomogeneous wet etching associated with the surface corrugation produced by substrate twinning.  Such behavior is not seen in regions of the sample where the twin domain blocks are parallel to the edge.  Atomic Force Microscopy (AFM) measurements show a one-to-one correspondence between the indented edges of the YBCO film and the LAO twin domain block steps (Fig. \ref{Fig4}).  The twin blocks have about 5 - 10 nm step heights.

	Fig. \ref{Fig3}(b) shows the simultaneously acquired rf photoresponse of this region of the film.  The edge-current enhancement is again clearly visible.  The rf photoresponse is stronger along edges of the film that have been more deeply etched.  The rf current path is clearly not straight along this edge.  Based on LSM and AFM images, the lateral variation in the current path is approximately 1 to 2 $\mu$m.  Fig. \ref{Fig3}(c) illustrates that besides being enhanced, the rf photoresponse is also distributed over a larger lateral length scale in regions where the etching has gone deeper into the film (e.g. location B).  The response is approximately 5 $\mu$m wider in those regions suggesting perhaps that the properties of the material (such as $\lambda$ and $\delta \lambda$) are larger for those parts of the film.

	Throughout this device we find a one-to-one correspondence between edges of the film disrupted by the underlying twinned substrate and significant enhancement of the rf photoresponse.  The enhanced rf photoresponse may be due to an increase in J$_{rf}^2$, or to an increase in $\lambda ^2$/t and $\delta \lambda$ at the position of the laser spot.  To demonstrate these latter types of contrast, we employed a second laser beam in the microscope (Fig. \ref{Schematic}), and focused it onto various parts of the film while imaging with the first beam.  The (fixed) location of the second beam shows up clearly in the image as increased photoresponse due to the additional local heating.  Thus the enhanced photoresponse of Fig. \ref{Fig3}(c) may be due to an increased effective penetration depth in that part of the film.

	The twinning of the LAO substrate clearly leads to steps and corrugation on the surface.  These features may create nano cracks \cite{Kastner,Koren} resulting in Josephson junctions and weak links in the overlying YBCO film.  These weak links will enhance the rf photoresponse because of the local increase in the effective penetration depth of the film. \cite{Hylton+Beasley}  Transverse averages of the rf photoresponse show $\sim$ 10-20\% increase in the deeply etched regions, compared to the other parts of the edge (see Fig. \ref{Fig3}(d)), suggesting $\sim$ 10\% increase in $\lambda$ there.

	Figure \ref{Fig4} also shows that the patterned edges of the film do not appear to be vertical.  The tapered edge of a superconducting transmission line has been identified as a microscopic source of enhanced nonlinearity.\cite{Dahm Edge}  Analysis of Fig. \ref{Fig4} shows that the edge angle is no less than 20$^0$, although this value is most likely limited by the bluntness of the AFM tip.  Hence one explanation for the wider distribution of rf photoresponse observed in Fig. \ref{Fig3}(c) may be that the more deeply etched edges of the film have a smaller edge angle, resulting in enhanced nonlinearity as well as an increase and broadening of rf photoresponse due to the smaller film thickness.

	We have also examined the temperature dependence of the photo-response distribution at one location.  We find that it is unchanged up to a temperature about 2 K below T$_c$, where $\lambda$(T)$^2$/2t is sufficiently large to see the current distribution visibly broaden.  This is consistent with the expected behavior of the current distribution. \cite{Brandt,van Duzer}  The rf power dependence of the photo-response was also investigated at one location, like position A in Fig. \ref{Fig3}.  No visible change in the rf photoresponse distribution was seen for input powers between -50 dBm and +10 dBm at 78 K, in agreement with Culbertson and Newman. \cite{C+N} We have also checked that the images are the same even with laser power levels 10 times smaller than those used for Figs. \ref{Fig2} and \ref{Fig3}.  

	We have developed a scanning laser microscope with a laser spot size of 1.5 $\mu$m, and a thermal spot size of about 4 $\mu$m and used it to measure the rf photoresponse of a superconducting microwave resonator.  We have established that the microscope is sensitive to the rf current density squared, and to the local penetration depth of the film.  The twinning of the LaAlO$_3$ substrate produces two types of irregularity in the rf photoresponse of the Ag-doped YBCO film.  First is a meandering of the current due to irregularities in the etching of the YBCO film associated with substrate twin domain blocks.  The second is an enhancement of the rf photoresponse at these locations, and this may be associated with enhanced inductance produced by weak links in the YBCO film or by a more shallow edge angle.  These irregularities are candidates for enhanced nonlinear response from the device, and will be investigated in detail in the future.

	We acknowledge the support of a NATO Collaborative Linkage Grant PST.CLG.977312, the Maryland Center for Superconductivity Research, the Maryland/Rutgers/NSF MRSEC DMR-0080008, and NSF/Neocera SBIR DMI-0078486.  We also thank K. Zaki for the design of the CPW resonator.

\begin{figure}[tbp]
\caption{Schematic diagram of LSM microscope.  The inset illustrates the principle of operation of the microscope.  The solid (dashed) line represents the transmission response of the unperturbed (perturbed) resonator.}
\label{Schematic}
\end{figure}

\begin{figure}[tbp]
\caption{The inset shows an LSM image of rf photoresponse over the length and breadth of the CPW strip conductor.  The image was obtained at 79.5 K at a frequency of 5.2133 GHz.  The figure shows a longitudinal (x) line cut (circles) of the image, taken along the upper edge of the YBCO strip.  The line cut and image are on the same horizontal scale.  The solid line is the fit to the photoresponse described in the text. For all figures, we take x to be the coordinate along the strip and y as the transverse coordinate.}
\label{Fig2}
\end{figure}

\begin{figure}[tbp]
\caption{Representative detailed view of the YBCO film on LAO substrate showing (a) reflectivity, and (b) photo-response of the surface in a 100 $\mu$m x 100 $\mu$m area.  c) shows transverse line cuts through the current distributions at locations A and B denoted in (b).  Note that the rf photoresponse is wider in location B compared to location A.  d) shows transverse averages of the data in (b), as a function of longitudinal position x, illustrating enhanced photoresponse in region B compared to region A.}
\label{Fig3}
\end{figure}

\begin{figure}[tbp]
\caption{AFM image of patterned edge of YBCO film (top) on LAO substrate (bottom).  The vertical dotted lines show the approximate locations of the LAO twin block domain step edges, while the "+" and "-" signs label the relative heights of the two regions.}
\label{Fig4}
\end{figure}

\end{document}